\documentclass[prb,twocolumn,superscriptaddress,showpacs,amsmath,amssymb]{revtex4-1}

\usepackage{graphicx}
\usepackage{latexsym}
\usepackage{amsmath}
\usepackage{amssymb}
\usepackage{amsfonts}
\usepackage{changes}
\usepackage{color}
\usepackage{bm}
\usepackage{verbatim}
\usepackage{pagecolor,lipsum}
\usepackage{lineno}
\usepackage[version=3]{mhchem}

\usepackage{framed}
\definecolor{shadecolor}{RGB}{222,222,221}
\begin{document}

\title{Long-range interaction of magnetic moments in a coupled system of S/F/S Josephson junctions with anomalous ground-state phase shift}

\author{G.A. Bobkov}
\affiliation{Moscow Institute of Physics and Technology, Dolgoprudny, 141700 Russia}

\author{I. V. Bobkova}
\affiliation{Institute of Solid State Physics, Chernogolovka, Moscow
  reg., 142432 Russia}
\affiliation{Moscow Institute of Physics and Technology, Dolgoprudny, 141700 Russia}
\affiliation{National Research University Higher School of Economics, Moscow, 101000 Russia}

\author{A. M. Bobkov}
\affiliation{Institute of Solid State Physics, Chernogolovka, Moscow reg., 142432 Russia}

\date{\today}


\begin{abstract}
A mechanism of a superconductivity-mediated interaction of two magnets in a system of coupled superconductor/ferromagnet/superconductor (S/F/S) Josephson junctions (JJs) with spin-orbit interaction  is proposed. The predicted indirect magnetic interaction favors the antiparallel orientation of the magnets. Its spatial scale is not restricted by the proximity length scales of the superconductor. Our estimates suggest that the interaction strength is not reduced considerably even at the macroscopic scales of the order of millimeters. At larger distances $l$ between the magnets the coupling constant exhibits the long-range power law $1/l$ behavior.  The mechanism of the interaction is based on two key ingredients: (i) the anomalous ground state phase shift in the S/F/S JJ provides a magnetoelectric coupling between the condensate phase and the magnetization  and (ii) the interaction is mediated by the condensate phase of the superconducting region connecting both JJs. In addition we demonstrate high tunability of the total magnetic configuration of the system by the externally controlled superconducting phase between the leads.  
\end{abstract}

  \maketitle
 
\section{Introduction}

Nowadays, heterostructures consisting of superconducting and magnetic materials are being actively studied both theoretically and experimentally \cite{Buzdin2005,Bergeret2005,Linder2015,Eschrig2015,Bergeret2018,Heikkila2019}. The reason for this interest is the possibility to realize in such hybrids properties and effects that are not possible in individual materials. In particular, one of the actively developing directions is the search and study of physical principles that can provide an indirect long-range interaction of magnetic moments through a superconductor. The indirect exchange interaction between magnetic moments carried by conduction electrons in a metal (RKKY interaction) is well known \cite{Ruderman1954}. It
has been studied in various materials \cite{Zhu2011,Abanin2011,Sherafati2011,Stefano2010,Hosseini2015,Liu2009}.  However, the strongly oscillating and decaying nature of this interaction at the atomic scale makes it possible to achieve interaction between magnetic moments at characteristic distances not exceeding a few nanometers in layered structures. 

In recent years experimental and theoretical studies, in which the nonmagnetic interlayer between the magnets in  spin valves is replaced by a superconductor \cite{Tagirov1999,Leksin2011,Li2013,DiBernardo2019,Ghanbari2021,Zhu2017,Koshelev2019,Aristov1997} have been actively carried out. As it was first pointed out by de Gennes, a superconductor makes the antiferromagnetic configuration of magnets more favorable \cite{deGennes1966}. The reason for this is that with such a mutual orientation of magnets, superconductivity in the interlayer is less suppressed as a result of partial compensation
of paramagnetic depairing. The characteristic scale of such an interaction is the superconducting coherence length $\xi_S$, at which the effect of proximity to a magnet manifests itself in a superconductor. It is tens to 
hundreds of nanometers, depending on the specific superconductor used. For the case of a $d$-wave superconductor the interaction length can be  enhanced due to the presence of nodal quasiparticles\cite{DiBernardo2019}. In recent work \cite{Devizorova2019} it was also proposed to use not the proximity effect to establish a coupling between magnets, but the so-called electromagnetic proximity effect \cite{Mironov2018}, the essence of which is the appearance of Meissner currents in a superconductor in response to the presence of an adjacent magnetic material.
The characteristic scale of this coupling is the penetration depth of the magnetic field. 

The interaction between localized magnetic moments through superconductors has also been studied \cite{Anderson1959,Galitski2002,Yao2014,Heimes2015,Zyuzin2014,Qin2014} and an additional to RKKY contribution decaying exponentially over $\xi_S$ and with a weaker power-law suppression, which favors an antiferromagnetic alignment, has been reported. Further, it has been shown \cite{Malshukov2018} that in superconductors with spin-orbit coupling (SOC) the superconducting condensate is coupled to the impurity spins, which results in more long-range non-exponential power-law suppression of the interaction between magnetic impurities. 

Here we propose a fundamental principle of using the superconducting state to establish (i) a total control over the magnetic configuration of two magnets, which are inserted into JJs and (ii) a long-range indirect interaction between their magnetic moments.  The interaction does not exploit proximity effects in superconductors and therefore, is not restricted by the typical proximity scales. The mechanism  is based on the fact that superconductivity is a macroscopic quantum state with a single phase of the condensate wave function and the condensate phase is coupled to the magnetization via the magnetoelectric effects. Then the ground state energy of a system of two coupled Josephson S/F/S junctions at a
given phase difference between the leads depends on the mutual orientation of the magnetizations of the ferromagnetic interlayers, which means an interaction between them.  The mechanism is of similar "magnetoelectric" origin as suggested in Ref.~\onlinecite{Malshukov2018} for impurity spins, but is realized in a very different class of physical systems.  

The effect can be observed in the systems, where a coupling between the direction of the magnetization of the magnet and the Josephson phase occurs. It is known that such a coupling physically manifests itself as the presence of an anomalous phase shift in the ground state of a Josephson junction and is realized in systems with a strong spin-orbit coupling \cite{Krive2004,Nesterov2016,Reynoso2008,Buzdin2008,Zazunov2009,Brunetti2013,Yokoyama2014,Bergeret2015,Campagnano2015,Konschelle2015,Kuzmanovski2016,Malshukov2010}. The strongest effect can be achieved in Josephson junctions on a
topological insulator \cite{Tanaka2009,Linder2010,Zyuzin2016,Lu2015,Dolcini2015}, because in these materials the coupling between the electron spin and its momentum is maximally
strong (spin-momentum locking) \cite{Burkov2010,Culcer2010,Yazyev2010,Li2014}. Josephson junctions with anomalous phase shift generated by the Zeeman effect of the applied magnetic field have already been implemented
experimentally by several groups \cite{Mayer2020,Szombati2016,Assouline2019,Murani2017}, including those on a topological insulator. Modern materials and techniques allow for realization of the anomalous ground state phase in S/F/S JJs. One of the possibilities is to use for the interlayers 2D or quasi 2D ferromagnets, where the Rashba spin-orbit coupling can be strong due to the structural inversion symmetry breaking. The other way is to exploit the ferromagnetic insulator/3D topological insulator (TI) hybrids as interlayers \cite{Chang2013,Kou2013,Kou2013_2,Chang2015,Jiang2014,Wei2013,Jiang2015,Jiang2016}. 

 \begin{figure}[h!]
 \centerline{$
 \begin{array}{c}
 \includegraphics[width=3.5in]{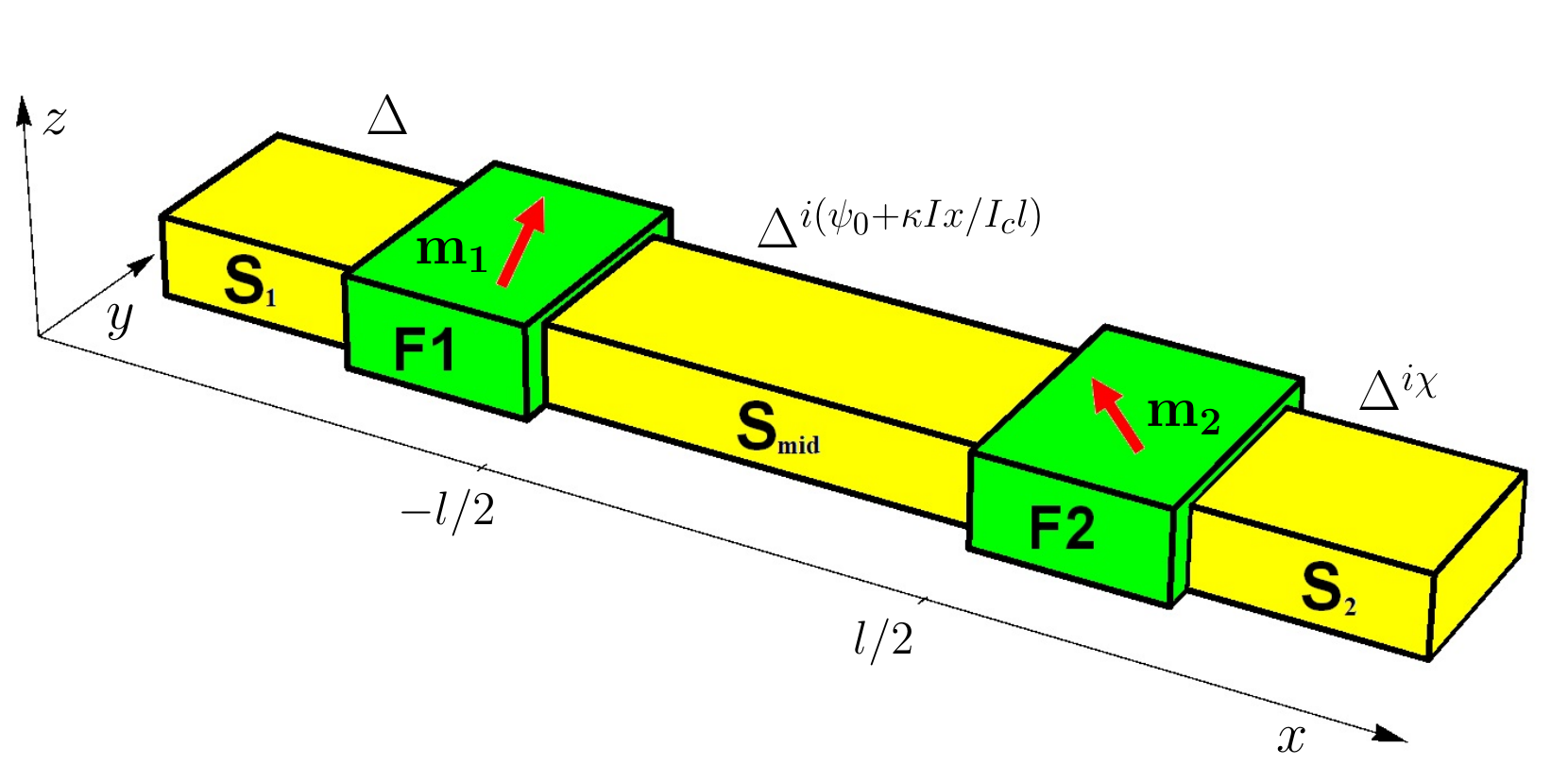} 
 \end{array}$}
 \caption{\label{sketch} 
Sketch of the coupled system of two S/F/S JJs. 
}
\label{sketch} 
 \end{figure}

\section{System and model} 

We consider two coupled S/F/S JJs, where S means a conventional superconductor and F means that the interlayer of each of the JJs consists of a spatially homogeneous ferromagnet with Rashba-type spin-orbit coupling. The spin-orbit coupling can be intrinsic or due to the structural inversion symmetry breaking, or it can be a hybrid interlayer consisting of a ferromagnet and a spin-orbit material, or it can be a ferromagnetic insulator on top of the 3D  TI. The last model is investigated in detail in the Appendix. If the ferromagnet is an insulator, it is assumed that the magnetization $\bm M$ of the ferromagnet induces an effective exchange field $\bm h \sim \bm M $ in the underlying conductive layer. The sketch of the system is represented in Fig.~\ref{sketch}. The superconducting phase difference $\chi$ between the leads is an external controlling parameter. First of all we investigate the energy of the system as a function of $\chi$ and $\bm m_1$ and $\bm m_2$, where $\bm m_i$ is the unit vector along the direction of the corresponding magnetization. It is assumed that the ferromagnets are easy-axis magnets with the easy axis along the $y$-direction. This choice of the easy-axis direction maximizes the magnetoelectric coupling between the magnetic moment and the superconducting phase, as it is discussed below.

The current-phase relation (CPR) of a separate S/F/S junction takes the form $I = I_c\sin (\chi_i-\chi_{0,i})$, where $\chi_{0,i}$ is the anomalous phase shift and $i=1,2$.  It has been found  that for Rashba-type SOC and ferromagnets on top of the 3D TI the anomalous phase shift is $\chi_0 =  r \hat {\bm  j} \cdot (\bm n \times \bm m )$, where $\hat {\bm  j}$ is the unit vector along the Josephson current and $\bm n$ is the unit vector describing the direction of the structural anisotropy in the system, in the case under consideration it is along the $z$-axis. The anomalous phase shift couples the superconducting phase to the magnetization direction. $r$ is a constant quantifying this coupling strength.  It is nonzero due to the presence of the Rashba SOC or the spin-momentum locking in the 3D TI surface states \cite{Linder2010,Zyuzin2016,Nashaat2019,Buzdin2008,Konschelle2015} and has been calculated in different models. For example, for Rashba-type SOC described by the hamiltonian $H_{R} = \alpha [\bm p \times \bm n] \bm \sigma$ ($\bm \sigma = (\sigma_x, \sigma_y, \sigma_z)^T$ is the vector of Pauli matrices) in the ballistic regime and for large Rashba constant $\alpha$, the constant $r$ is given by \cite{Buzdin2008} 
\begin{eqnarray}
r_{b} = \frac{4 h \alpha d}{(\hbar v_F)^2} ,
\label{phi0_ballistic}
\end{eqnarray}
where $d$ is the length of the Josephson junction interlayer and $v_F$ is the Fermi velocity of the electrons in the interlayer, $h$ is the absolute value of the exchange field in the interlayer of the JJ. In the diffusive regime for weak $\alpha$, highly transparent interfaces and neglecting spin-relaxation, the predicted result for the constant $r$ is
\begin{eqnarray}
r_{d} = \frac{\tau m^{*2} h (\alpha d)^3}{3 \hbar^6 D} ,
\label{phi0_diffusive}
\end{eqnarray}
where $\tau$ is the elastic scattering time, $m^*$ is the effective electron mass and $D$ is the diffusion constant \cite{Bergeret2015}. For the S/F/S JJs on top of the 3D TI it has been predicted that $r = 2hd/v_F$ \cite{Zyuzin2016,Nashaat2019}. 
Therefore, we can conclude that symmetry of our system dictates that
\begin{eqnarray}
\chi_{0,i} = r m_{yi}
\label{chi0}
\end{eqnarray}
irrespective of the particular model. This relation also survives in the dynamic situation $\bm m_i = \bm m_i(t)$ and has been used  for calculation of the magnetization dynamics in voltage-biased and current-biased JJs \cite{Nashaat2019,Konschelle2009,Shukrinov2017,Guarcello2020}.  

The  critical current depends crucially on the particular model. For example, it can be independent on the magnetization direction, as it has been reported for the ferromagnets with SOC \cite{Buzdin2008}, or it can depend strongly on  the $x$-component of the magnetization, as it takes place for the ferromagnetic interlayers on top of the 3D TI \cite{Zyuzin2016,Nashaat2019}. Here we focus on the model where $I_c$ does not depend on the magnetization direction. The influence of the dependence $I_c(\bm m)$ on the results is considered in detail in the Appendix. The energy of the system consists of the Josephson energies of  both junctions and the easy-axis anisotropy energies of  both magnets:
\begin{eqnarray}
E = \frac{\hbar}{2e}\Bigl[ I_{c}\bigl(1-\cos(\psi_1-\chi_{0,1})\bigr) +~~~~~~~~~ \nonumber \\
I_{c}\bigl(1-\cos(\chi-\psi_2-\chi_{0,2})\bigr)\Bigr]- 
\frac{KV_F}{2}(m_{y1}^2+m_{y2}^2),
\label{energy}
\end{eqnarray}
where $K$ - is the anisotropy constant, $V_F$ is the volume of the ferromagnet.
$\psi_{1,2}$ are the values of  the phase of the middle superconductor ($\rm S_{mid}$ in Fig.~\ref{sketch}) at the ${\rm F1/S_{mid}}$ and ${\rm S_{mid}/F2}$ interfaces. $\psi_{1,2} = \psi_0 \mp \kappa (I/2I_c)$, where the second term accounts for the phase gradient due to the supercurrent flowing through the system and $\kappa \propto l$, where $l$ is the length of $\rm S_{mid}$. The current conservation dictates 
\begin{eqnarray}
I_{c}\sin(\psi_1-\chi_{0,1}) = 
I_{c}\sin(\chi-\psi_2-\chi_{0,2}) .
\label{current_conservation}
\end{eqnarray}
Eliminating the phase $\psi_0$  making use of Eq.~(\ref{current_conservation}), the energy of the coupled JJs takes the form:
\begin{eqnarray}
E = 2 E_J \bigl[1-\cos(\frac{\chi}{2}-\bar \chi_0  - \kappa\frac{I}{2I_c} + \pi n) \bigr]- \nonumber \\
E_M (m_{y1}^2+m_{y2}^2), 
\label{energy_2}
\end{eqnarray}
where $E_J = \hbar I_c/2e$, $E_M = K V_F/2$, $\bar \chi_0 = (\chi_{0,1}+\chi_{0,2})/2$ and $n$ is an integer number.
Eq.~(\ref{energy_2}) should be supplied by the "self-consistency equation" for the Josephson current:
\begin{eqnarray}
I = I_c \sin (\frac{\chi}{2}-\bar \chi_0  - \kappa\frac{I}{2I_c} + \pi n) .
\label{self_consist}
\end{eqnarray}

\section{Phase-dependent stable magnetic configurations} 

At first we discuss the dependence of the  total magnetic configuration $(\bm m_1, \bm m_2)$ on the external phase difference $\chi$ and its tunability by this parameter. To simplify the analysis we disregard the order parameter phase gradient $\kappa$ in the middle superconductor. As it is suggested by our estimates of $\kappa$ (see below), this approximation should be valid up to the submillimeter scale. The influence of $\kappa$ on the magnetic configuration and its tunability is discussed at the end of this section.

\begin{figure*}[tbh!]
 \centerline{$
 \begin{array}{c}
 \includegraphics[width=6.6in]{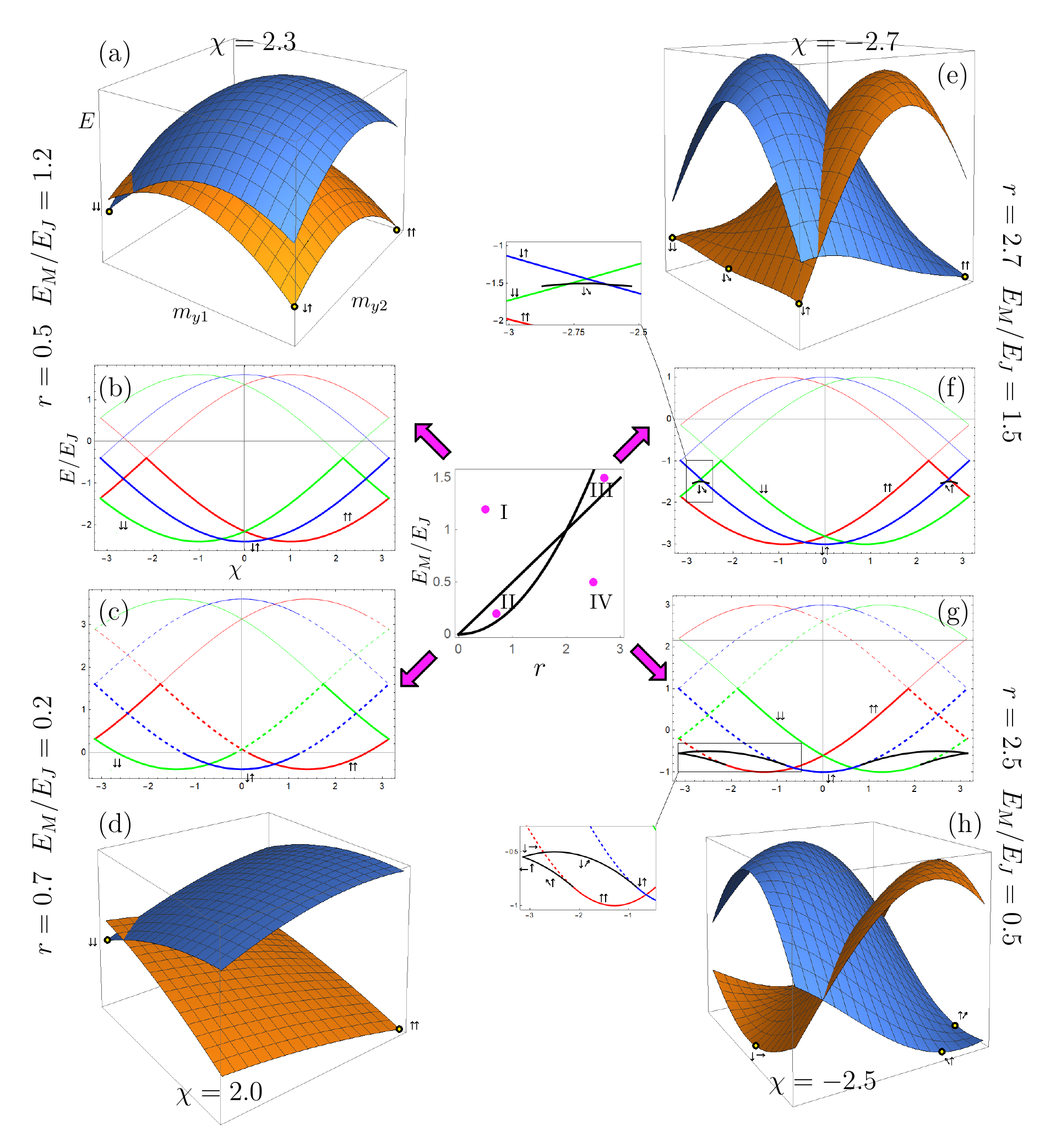} 
 \end{array}$}
 \caption{Middle: the phase diagram representing four physically different regions corresponding to the "corner" stable/metastable magnetic configuration (I); "corner" stable/unstable (II); "corner"+"non-aligned" stable/metastable (III) and "corner"+"non-aligned" stable/unstable (IV) extremum states. (a) Energy of the system for the pink point from region I as a function of $(m_{y,1},m_{y,2})$ at $\chi = 2.3$. (b) Energy branches of the corner magnetic configurations for the pink point from region I as functions of $\chi$. The thin part of each branch is unstable with respect to a $\pi$-shift of $\psi_0$ and jump to the corresponding bold part. (c)-(d) The same as (a)-(b) but for the pink point from region II. The dashed parts of the branches correspond to unstable corner states. (e)-(f) the same as (a)-(b) but for the pink point from region III. (g)-(h) the same as (a)-(b) but for the pink point from region IV. $\kappa = 0$.}
\label{config} 
\end{figure*}

Neglecting $\kappa$ Eq.~(\ref{energy_2}) is reduced to:
\begin{eqnarray}
E_{\pm} = 2 E_J\bigl[1\mp\cos(\frac{\chi}{2}- \nonumber \\ \frac{r(m_{y1}+m_{y2})}{2})\bigr]- 
E_M(m_{y1}^2+m_{y2}^2) .
\label{energy2}
\end{eqnarray}
The energy $E$ as a function of $(m_{y1},m_{y2})$ at a given $\chi$ consists of two branches $E_\pm$, which differ by the phase $\pi$ at ${\rm S_{mid}}$. Examples of the corresponding plots are presented in Figs.~\ref{config}(a),(d),(e) and (h). The upper energy value at a given magnetic configuration is unstable.  Now we focus on the magnetic configurations, corresponding to the extrema of the energy. The magnetic part of the energy has a minimum at $m_{y1(2)} = \pm 1$. We call the states with $m_{y1} = \pm 1$ and $m_{y2} = \pm 1$ by the "corner states". Let us consider the energy in the vicinity of $m_{y1} = m_{y2} = 1$.  At $rE_J/2  < E_M$ this corner point is always a minimum of the energy Eq.~(\ref{energy2}) at any phase difference. On the contrary, at $rE_J /2 > E_M$ it can become a maximum of the energy Eq.~(\ref{energy2}) at a particular value of $\chi$. The situations corresponding to the other "corner states" lead to the same result. Consequently, the corresponding magnetic configuration can be made absolutely unstable by varying the phase. Thus, the parameter $r$ removes the degeneracy between the "corner" states, making some of them stable and the others unstable at a given phase difference. It allows for the control of the magnetic configuration by variations of the superconducting phase $\chi$. The other important parameter in the system is the ratio of the magnetic anisotropy and Josephson energies $E_M/E_J $. The larger the parameter $E_M/E_J$ the higher the energy barrier between the different stable states, which worsens the tunability. 

The condition $E_M/E_J < (r/2)^2$ allows for appearance of additional minima of the energy Eq.~(\ref{energy2}), which differ from the corner states. Indeed, in order to have an energy minimum at $m_{y1} \neq \pm 1$, we need $\partial E_\pm /\partial m_{y1}=0$ and $\partial^2 E_\pm /\partial m_{y1}^2 = \pm 2 (r/2)^2 E_J \cos [\chi/2 - r(m_{y1}+m_{y2})/2] - 2 E_M>0$, which is only possible under the above condition. The minima corresponding to $m_{y1} \neq 1$ and $m_{y2} \neq 1$ do not occur in this model.

The two lines $E_M/E_J = r/2$ and  $E_M/E_J = (r/2)^2$ divide the phase diagram of the system into four regions, which are marked by numbers I-IV in the central panel of Fig.~\ref{config}. As discussed above, in regions I and III all the corner magnetic configurations are stable/metastable at an arbitrary phase difference. It leads to the absence of unstable parts of the energy branches in Figs.~\ref{config}(b) and (f). At the same time the analogous figures (c) and (g), corresponding to regions II and IV, respectively, have unstable parts. It means that in these regions the magnetic configuration can be easily manipulated by the phase variations. 

The right column of Fig.~\ref{config} corresponds to regions III and IV, where "non-aligned" stable magnetic states are possible at particular values of $\chi$. The "non-aligned" stable states are demonstrated in Figs.~\ref{config}(e) and (h) by points and arrows. The energy of these stable "non-aligned" states as a function of  $\chi$ is represented in Figs.~\ref{config}(f) and (g) by black lines. The ranges of $\chi$ values, where the "non-aligned" stable states exist, are small and for this reason the corresponding parts of the energy branches are shown on larger scale. Therefore, in region IV the magnetic configuration of the system can be switched between $\uparrow \uparrow$, $\downarrow \downarrow$, AP and "non-aligned" states by varying the phase difference. With good accuracy the influence of nonzero $\kappa$ on the phase diagram can be taken into account by replacing $r \to r/(1+\kappa/2)$.

We concentrate on the parameters falling into region II of the phase diagram. On the one hand, in this region the magnetic configuration is tunable by the phase difference. On the other hand, the physical picture is more transparent here because of the absence of the non-aligned extreme states. Therefore, all the extreme magnetic configurations are realized by the corner states. An example of the energy of the corner magnetic configurations as a function of $\chi$ is presented in Fig.~\ref{config}(c). Due to the reflection symmetry with respect to the $(x,z)$-plane  $E_{\uparrow \uparrow}(\chi) = E_{\downarrow \downarrow}(-\chi)$ and $E_{\uparrow \downarrow}(\chi) = E_{\downarrow \uparrow}(-\chi)$. It is seen from Fig.~\ref{config}(c) that at  $r \neq 0$ $E_{\uparrow \uparrow}(\chi)$ and $E_{\downarrow \downarrow}(\chi)$ are asymmetric functions of $\chi$ and, therefore, the degeneracy between them is removed. At the same time $E_{\uparrow \downarrow}(\chi)$ is a symmetric function of $\chi$ and, consequently, the states $\uparrow \downarrow$ and $\downarrow \uparrow$ remain degenerate and we refer to them as the antiparallel (AP) state. $\uparrow \uparrow$, $\downarrow \downarrow$ and AP states can be stable (solid) or unstable (dashed) depending on $\chi$. The upper branches of the energy, which are unstable with respect to the $\pi$-jump of ${\rm S_{mid}}$ phase, are shown by thin lines. Each of the states represents the ground state of the system for the particular range of $\chi$. Thus, at the chosen parameters any of the corner magnetic states can be realized by adjusting the phase, that is the total control over the magnetic configuration is possible.

We have estimated the parameters $E_M/E_J$ and $r$ for the model of the insulating ferromagnet on top of the 3D TI. We take the parameters corresponding to $Nb/Bi_2Te_3/Nb$ Josephson junctions \cite{Veldhorst2012}: the junction length $d=50 nm$, $I_{c} = 40 A/m$, $v_F = 10^5 m/s$. We assume $E_M \sim [(10-10^2)erg/cm^3]\times d_F$  for YIG thin films \cite{Mendil2019}, where $d_F = 10 nm$ is the F thickness along the $z$-direction. It gives $E_M/E_J \sim 10^{-2}-10^{-1}$. Basing on the experimental data on the Curie temperature of the magnetized TI surface states \cite{Jiang2015}, where the Curie temperature in the range $20-150K$ was reported, we can roughly estimate $h \lesssim 0.01- 0.1 h_{YIG}$. It corresponds to the dimensionless parameter $r = 2 h d/v_F \lesssim 2-13$.

\section{Long-range indirect magnetic interaction.} 

The minima of all the energy branches, see Fig.~\ref{config}(c), correspond to $I=0$. In the vicinity of the minima $E$ can be approximated by $E = E_J (I/I_c)^2 - E_M(m_{y1}^2+m_{y2}^2).$ The current $I$ can be found from Eq.~(\ref{self_consist}) as $I/I_c \approx \chi/2 - \bar \chi_0 - \kappa (I/2I_c) + \pi n$ and $n$ is chosen to have $I/I_c $ close to zero, which results in 
\begin{eqnarray}
\frac{I}{I_c} = \frac{\tilde \chi - 2\bar \chi_0}{2+\kappa}, 
\label{current_3}
\end{eqnarray} 
where $\tilde \chi = \chi + 2\pi n$. Substituting Eq.~(\ref{current_3}) into the energy, we obtain 
\begin{eqnarray}
E \approx - \frac{2 E_J  r(m_{y1}+m_{y2})\tilde \chi}{(2+\kappa)^2}+\frac{E_Jr^2(m_{y1}^2+m_{y2}^2)}{(2+\kappa)^2} + \nonumber \\
  \frac{2E_J r^2 m_{y1}m_{y2}}{(2+\kappa)^2} - E_M(m_{y1}^2+m_{y2}^2) + const. ~~~~~~~~~  
\label{energy_3}
\end{eqnarray}
The first term in Eq.~(\ref{energy_3}) accounts for the individual coupling of the magnetic moments to the phase, the second term works as an additional contribution to the magnetic anisotropy and the third term describes the interaction between the moments. The coupling constant $J_{eff} = 2E_J r^2/(2+\kappa)^2>0$ and, therefore, the interaction favors the antiparallel alignment. At larger external phases $\chi$ the first term dominates resulting in the $\uparrow \uparrow$ or $\downarrow \downarrow$ ground state, as it is seen from Fig.~\ref{config}(c), but at smaller phases the antiferromagnetic interaction overcomes this term. The spatial dependence of $J_{eff}$ is determined by $\kappa \propto l$, that is $J_{eff} \propto l^{-2}$ for large $l$. For estimates of $\kappa \sim e I_c l/\sigma_S \Delta S$, where $\Delta$ is the superconducting order parameter, $\sigma_S$ is the normal state conductivity of the middle superconductor and $S$ is its cross section, we take typical parameters of $Nb/Bi_2Te_3/Nb$ JJs \cite{Veldhorst2012} $I_c = w \cdot [40 A/m]$, where $w \sim 1\mu m$ is the width of the JJ along the $y$-direction, $\sigma_S = \sigma_{Nb} = 10^7 (\Omega \cdot m)^{-1}$, $\Delta_{Nb} = 2.5 \cdot 10^{-22}$J and  $S=(1\mu m)^2$. Then $\kappa \sim 1$ at $l \sim 1 mm$. Therefore, the results represented in Fig.~\ref{config} and calculated at $\kappa = 0$ are applicable for the distances between the magnets up to submillimeter  scale. In addition, for the lengths of the middle superconductor of the order of $\sim 1 mm$ the inductance energy $E_L = LI^2/2$ becomes of the same order of magnitude as the Josephson energy and should be taken into account. Accounting for this energy results in the substitution $E_J \to E_J + LI_c^2/2$ in Eq.~(\ref{energy_3}). Because of $L \propto l$ that modifies $J_{eff} \propto l^{-1}$ at large distances between the magnets.

 \begin{figure}[h!]
 \centerline{$
 \begin{array}{c}
 \includegraphics[width=3.6in]{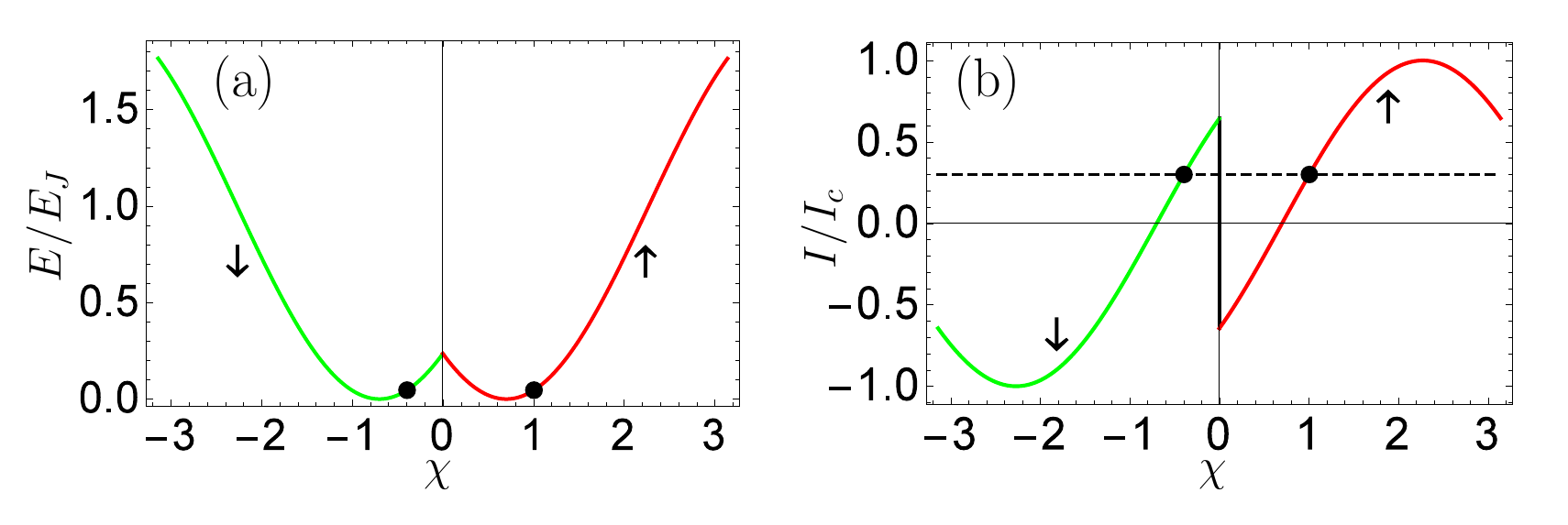} 
 \end{array}$}
 \caption{(a) Ground-state energy of the individual S/F/S junction with the anomalous phase shift. Due to the presence of the anomalous phase shift, which depends on the magnetization direction, the standard cosine energy curves for  both magnetization orientations are shifted. As a result the ground state at a given phase is realized by different magnetization orientations, as shown by different colors and arrows in the figure. (b) CPR of the individual S/F/S junction with the anomalous phase shift. For a given value of the applied current, shown by the dashed line, there are two states of the system supporting the current. They are marked by the black points and correspond to the opposite orientations of the magnetic moment, but are degenerate in energy, as it is indicated by the same black points in panel (a).}
\label{current_interaction} 
\end{figure}

It is worth to stress that the effective magnetic interaction described by Eq.~(\ref{energy_3}) can be only realized under the fixed superconducting phase $\chi$. If instead we consider the current $I$ as an external fixed parameter, there is no interaction between the magnets. The reason is explained in Fig.~\ref{current_interaction}, where the ground-state energy of the individual S/F/S junction with the anomalous phase shift and the CPR of the junction are presented as functions of $\chi$.  It is seen that if we fix a current [dashed line in Fig.~\ref{current_interaction}(b)], then for not very large current values this current can be supplied by two different phases $\chi$. Both values of $\chi$ describe energetically degenerate states, as it is demonstrated in Fig.~\ref{current_interaction}(a), but correspond to the opposite directions of the magnetic moment of the interlayer. Therefore, at small enough applied currents the orientation of each of the magnets is chosen by the system spontaneously and independently on the orientation of the other magnet. 

Experimentally the phase $\chi$ can be controlled by several ways. One of them is to insert the considered system into the superconducting loop under the applied magnetic flux, the other way is to insert it into the asymmetric Josephson interferometer, where the considered system is in parallel with an ordinary Josephson junction with a much higher critical current. Then the magnetic state of the system can be controlled by the external current. Moreover, if the system is in the regime of the AP ground state, where the interaction between the magnets dominates over the individual interactions of the magnets with the phase, the orientation of a magnet can be remotely switched by the  external impact on the other magnet. Further we investigate the dynamics of the above mentioned processes.

\section{Dynamics}  

The dynamics of each of the magnets $i=1,2$ is described by the Landau-Lifshitz-Gilbert (LLG) equation: 
\begin{eqnarray}
\frac{\partial\bm m_i}{\partial t} = -\gamma \bm m_i \times \bm H_{eff} + \alpha \bm m_i \times \frac{\partial\bm m_i}{\partial t} - \nonumber \\
\frac{\gamma r I}{2e M d d_F}[\bm m \times \bm e_y],~~~~~~
\label{LLG}
\end{eqnarray}
where $\gamma$ is the gyromagnetic ratio,  $\bm H_{eff} = (K/M) m_y \bm e_y$ is the local effective field in the ferromagnet induced by the easy-axis magnetic anisotropy and $\alpha$ is the Gilbert damping constant. The last term in Eq.~(\ref{LLG}) describes the spin-orbit torque, exerted on the magnet by the electric current $I$  \cite{Yokoyama2011,Miron2010,Bobkova2018,Bobkova2020}. The torque is averaged over the ferromagnet thickness $d_F$ along the $z$-direction. 
The total current flowing through each of the JJs consists of the supercurrent and the normal quasiparticle current contributions \cite{Rabinovich2019,Rabinovich2020}:
\begin{eqnarray}
I=I_c \sin (\chi_i - \chi_{0,i})+\frac{1}{2eR_N}(\dot \chi_i - \dot \chi_{0,i}),
\label{current_total}
\end{eqnarray}
where $\chi_{1} = \psi_0(t)$ and $\chi_2 = \chi(t)-\psi_0(t)$. Here we assume $\kappa = 0$. The dynamics of the magnetizations $\bm m_{1,2}$ is calculated numerically from Eqs.~(\ref{LLG}) and (\ref{current_total}). The equations for the both JJs are coupled via the phase $\psi_0(t)$. 
If the normal current, represented by the second term in Eq.~(\ref{current_total}) is small, the torque is mainly determined by the supercurrent and can be calculated via the additional contribution to the effective field in Eq.~(\ref{LLG}) $\delta \bm H_{eff} = -(1/M d d_F)dE/d\bm m$ \cite{Nashaat2019,Konschelle2009,Shukrinov2017,Guarcello2020}, which leads to Eq.~(\ref{LLG}) with $I \to I_c \sin(\chi_i-\chi_{0,i})$.

 \begin{figure}[h!]
 \centerline{$
 \begin{array}{c}
 \includegraphics[width=3.6in]{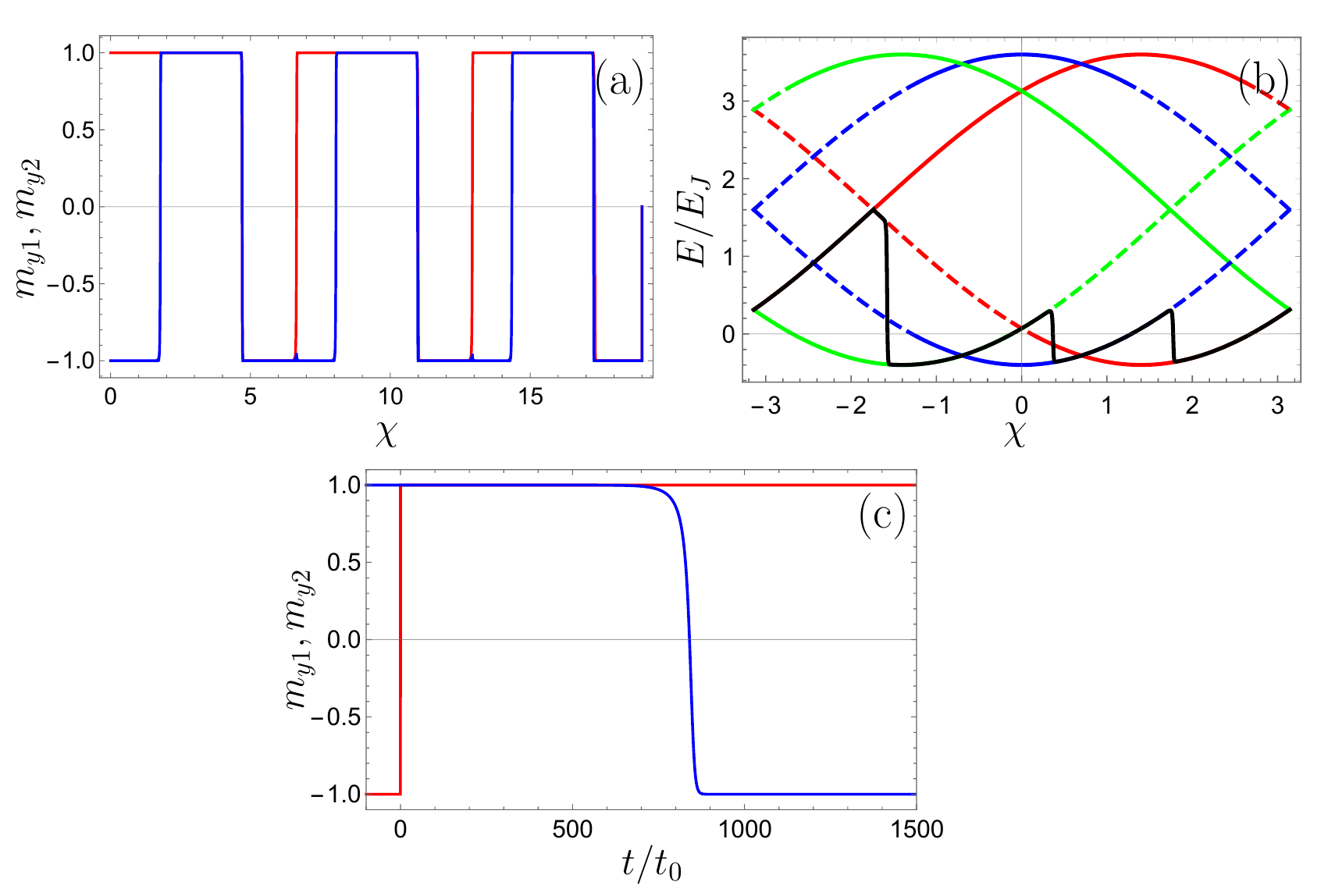} 
 \end{array}$}
 \caption{(a) Time evolution of $m_{y1}$ (red) and $m_{y2}$ (blue) under the adiabatic variation of the phase $\chi= 2eV t$. $t_0=M/\gamma K$, $eVt_0 =10^{-4}$, $2eR_N I_c t_0 = 10^3$,  $r=0.7$, $E_M/E_J =0.2$, $\alpha = 0.02$, $\kappa = 0$. (b) Matching the dynamic magnetic configuration (black line), presented in (a) to the energy of the system. (c) Time evolution of $m_{y1}$ (red) and $m_{y2}$ (blue) initiated by $\bm m_1$  reversal at $t=0$ under a given $\chi$. }
\label{adiabatic} 
\end{figure}

The resulting control over the magnetic state of the system (corresponding to the parameters falling into region II of the phase diagram) by the adiabatic phase variation is demonstrated in Fig.~\ref{adiabatic}(a)-(b). Fig.~\ref{adiabatic}(a) represents  $m_{y1}$ (red) and $m_{y2}$ (blue) as functions of $\chi \propto t$ starting from the initial AP configuration. In Fig.~\ref{adiabatic}(b) we  match the dynamic magnetic configuration of the system with the energy of the equilibrium state at the same phase difference. The matching is performed for the phase interval $\chi \in (\pi, 3\pi)$.  

The results of the remote switching of $\bm m_2$ by the external impact on $\bm m_1$ are demonstrated in Fig.~\ref{adiabatic}(c). The phase $\chi$ is chosen in such a way that the equilibrium magnetic configuration is AP. At $t=0$ $\bm m_1$ is fixed in the new position by external means (for example, by the applied magnetic field). It is seen that $\bm m_2$ also switches in order to make the magnetic configuration AP, which is energetically favorable at the given $\chi$. The characteristic time of the reversal is  much larger than $t_0 = M/\gamma K$, which is the characteristic time of magnetization dynamics, and depends essentially on the particular value of $\chi$. 

\begin{figure*}[tbh!]
 \centerline{$
 \begin{array}{c}
 \includegraphics[width=7.2in]{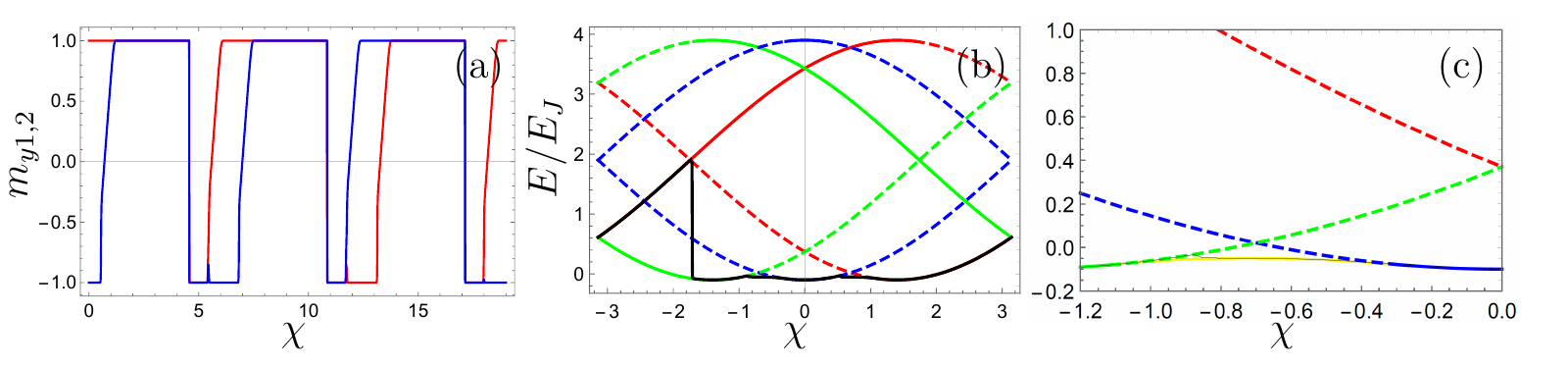} 
 \end{array}$}
 \caption{(a) Time evolution of $m_{y1,2}$ under the adiabatic variation of the phase $\chi= 2eV t$ in region IV ($r=0.7$, $E_M/E_J =0.05$). $eVt_0 =5\times 10^{-4}$. (b) Matching the dynamical magnetic configuration, presented in (a) to the energy of the system. Black line is the dynamical trajectory of the system. (c) Energy in the range $\chi \in (-1.2,0)$, where the system switches from $\downarrow \downarrow$ to AP configuration via the non-aligned state, on a larger scale. The yellow line is the equilibrium non-aligned energy branch and the thin black line is the dynamical trajectory of the system. }
\label{s2} 
\end{figure*}

The dynamics of the magnetic configuration in region IV of the phase diagram under the adiabatic phase variations $\chi = 2eVt$ is shown in Fig.~\ref{s2}. Fig.~\ref{s2}(a) demonstrates that the switching between $\downarrow \downarrow$ and AP configurations occurs via the non-aligned states, where one of the $y$-components of the magnetization is less than unity. Fig.~\ref{s2}(b) illustrates matching between the dynamical trajectory of the time evolution of the magnetic configuration and the equilibrium energy branches. 

\section{Conclusions}

In conclusion, we have proposed a mechanism of long-range antiferromagnetic interaction via the superconducting phase between the magnets incorporated into a system of coupled S/F/S JJs. It is based on (i) the magnetoelectric coupling between the condensate phase difference and the magnetization in the weak link of the JJs with anomalous ground state phase and (ii) the macroscopic character of the superconducting phase in the middle superconductor, which interacts with both magnets thus mediating the interaction between them. The interaction strength is not determined by the proximity length scales and decays $\propto l^{-1}$ at large distance $l$ between the magnets. It is also demonstrated that the total magnetic configuration of the system can be controlled and manipulated via the superconducting phase.

\begin{acknowledgments}
The work of I.V.B and A.M.B has been carried out within the state task of ISSP RAS. The numerical analysis of the dynamics has been supported by RSF project No. 18-72-10135. I.V.B. also acknowledges the financial support by Foundation for the Advancement of Theoretical Physics and Mathematics “BASIS”.
\end{acknowledgments}

\section{Appendix: Role of the dependence of the critical current on the magnetization direction.}

\begin{figure*}[tbh!]
 \centerline{$
 \begin{array}{c}
 \includegraphics[width=7.2in]{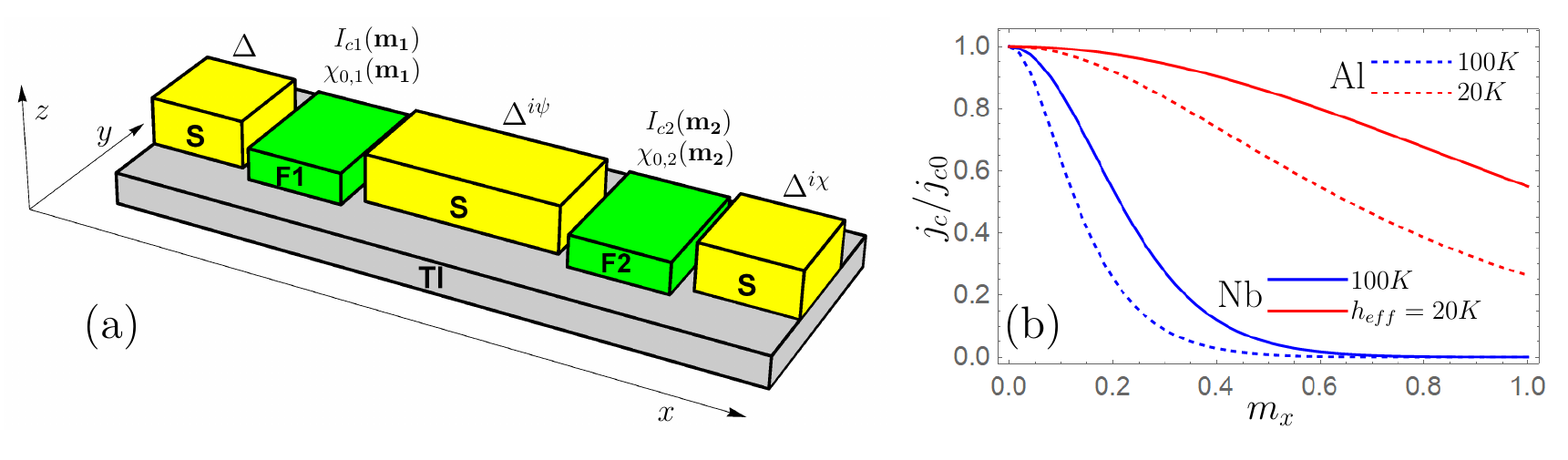} 
 \end{array}$}
\caption{(a) Sketch of the system of two coupled S/F/S JJs on top of a 3D TI. (b) $I_c$ as a function of $m_x$ for $r=13.2$, $d/\xi_N =4.1 $ (solid blue); $r=2.6$, $d/\xi_N = 4.1$ (solid red);  $r=13.2$, $d/\xi_N =0.74 $ (dashed blue); $r=2.6$, $d/\xi_N =0.74 $ (dashed red). $I_c$ is normalized to $ I_c(m_x=0)$.}
\label{s3} 
\end{figure*}

Here by considering the particular model of the S/F/S JJs on top of the 3D TI we investigate the role of the dependence of the critical current on the magnetization direction $I_c(\bm m)$. The  interlayer region of a S/3D TI/S JJ is covered by a ferromagnet. We believe that our results can be of potential interest for systems based on $Be_2Se_3/YIG$ or $Be_2Se_3/EuS$ hybrids, which were realized experimentally. It is assumed that the ferromagnet induces an effective exchange field $\bm h \propto \bm M$ (where $\bm M$ is the ferromagnet magnetization) in the underlying 3D TI surface states, as it has been reported experimentally \cite{Jiang2015}. The sketch of the setup is shown in Fig.~\ref{s3}(a).

The Josephson current assuming the ballistic limit for the 3D TI surface states and in the vicinity of the critical temperature takes the form \cite{Nashaat2019}:
\begin{eqnarray}
I_s = I_c \sin (\chi - \chi_0),~~~~~~~~~~~~~~ \label{Josephson_CPR}\\
I_c = I_b \int \limits_{-\pi/2}^{\pi/2} d \phi \cos \phi  \exp[-\frac{2\pi T d}{v_F \cos \phi}] \cos [r m_x \tan \phi],~~~~  \label{critical_current} \\
\chi_0 = 2 h_y d/v_F = r m_y, ~~~~~~~~~~~~~
\label{chi_0}
\end{eqnarray}
where $r = 2hd/v_F$ for the 3D TI and $I_b = ev_F N_F \Delta^2/(\pi^2 T)$, $v_F$ and $N_F$ are the Fermi velocity and the normal state density of states at the 3D TI surface. Here the critical Josephson current is only suppressed by the $x$-component of the exchange field. The $y$-component of the field does not lead to the suppression, instead it gives rise to the anomalous phase shift. This statement is also valid for the diffusive case. The Josephson current in 3D TI-based Josephson current has been considered in Ref.~\onlinecite{Zyuzin2016} and exactly the same expression for the anomalous phase shift $\chi_0$ has been obtained. The result for the critical current is different in the diffusive case, but it still only depends on the $x$-component of the exchange field. The suppression of the critical current as a function of $m_x \equiv M_x/M_s$ is presented in Fig.~\ref{s3}(b). For estimates we take  $d=50nm$, $v_F = 10^5m/s$  and $T_c = 10K$, which corresponds to the parameters of $Nb/Bi_2Te_3/Nb$ Josephson junctions\cite{Veldhorst2012}. In this case $\xi_N = v_F/2\pi T_c \approx 12nm$. We have also plotted $I_c(m_x)$ for $T_c = 1.8K$, what corresponds to the Josephson junctions with $Al$ leads.

\begin{figure*}[tbh!]
 \centerline{$
 \begin{array}{c}
 \includegraphics[width=5.5in]{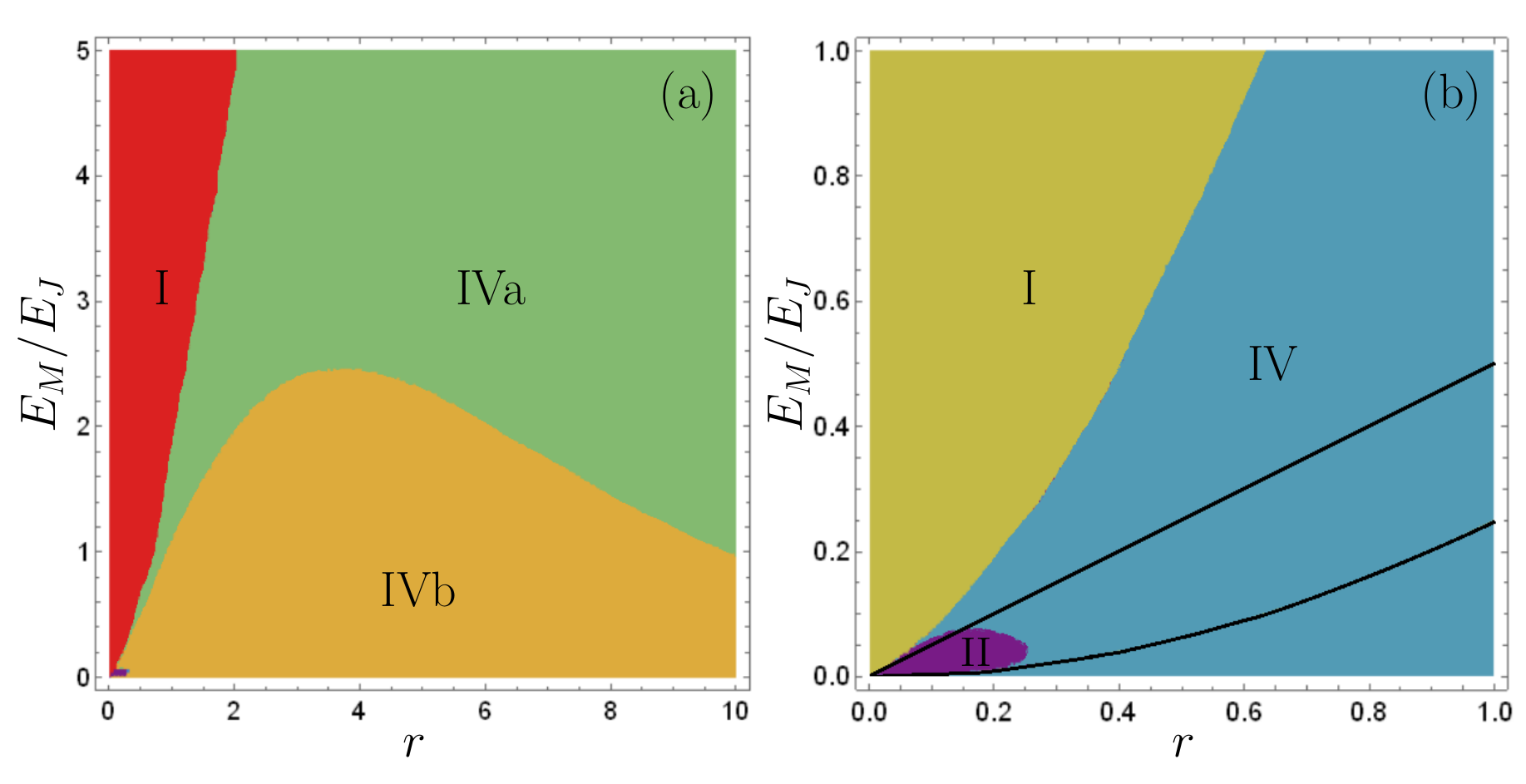} 
 \end{array}$}
 \caption{(a) Phase diagram of the S/F/S JJ on top of the 3D TI. For description of the different regions see text. (b) The bottom left corner of the diagram on a large scale.}
\label{s4} 
\end{figure*}

\begin{figure*}[tbh!]
 \centerline{$
 \begin{array}{c}
 \includegraphics[width=5.5in]{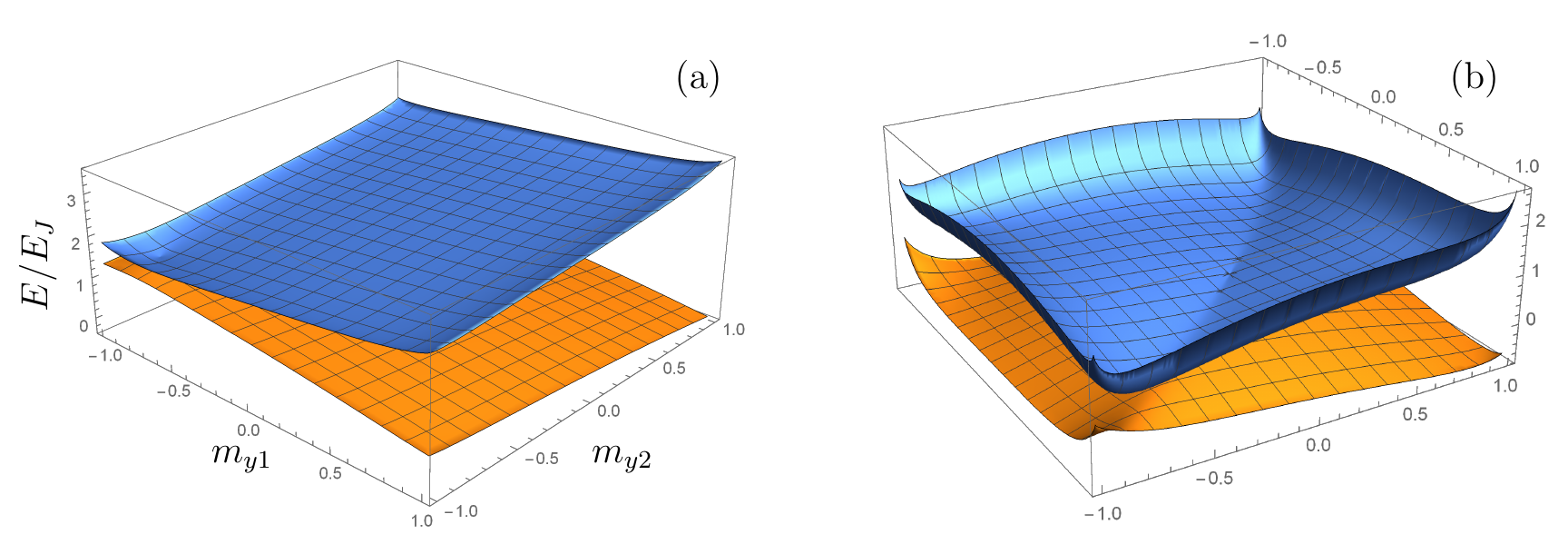} 
 \end{array}$}
 \caption{Energy as function of $(m_{y1},m_{y2})$. (a) $\chi=2.0$, $r=0.7$ and $E_M/E_J = 0.2$; (b) $\chi=-2.5$, $r=2.5$ and $E_M/E_J = 0.5$.}
\label{s5} 
\end{figure*}

Making use of the current conservation condition Eq.~(\ref{current_conservation}), the energy of the system can be expressed in the form:
\begin{eqnarray}
E_\pm = \frac{\hbar}{2e}\Bigl[I_{c1}(m_{x1}) + I_{c2}(m_{x2})\mp I(\bm m_1, \bm m_2)\Bigr]- \nonumber \\ 
\frac{KV_F}{2}(m_{y1}^2+m_{y2}^2),~~~~~~~~~~~
\label{energy3}
\end{eqnarray}
where 
\begin{eqnarray}
I(\bm m_1, \bm m_2) = 
\sqrt{I_{c1}^2 + I_{c2}^2 +  2 I_{c1} I_{c2}\cos (\chi - \chi_{0,1} - \chi_{0,2})}~~~~~~~
\label{current_mx}
\end{eqnarray}
and $I_{c1(2)} = I_{c1(2)}(m_{x1(2)})$. 
Eq.~(\ref{energy3}) is exploited to calculate the phase diagrams, presented in Fig.~\ref{s4} and the energy surfaces in Fig.~\ref{s5}. Physically different regions of the phase diagram are marked by the same numbers as for the previous model with constant critical current. Fig.~\ref{s4}(b) is the bottom left corner of the phase diagram, presented in Fig.~\ref{s4}(a) on a larger scale. It demonstrates region II, which is very small in Fig.~\ref{s4}(a). The black curves in this figure represent the lines  $E_M/E_J = r/2$ and  $E_M/E_J = (r/2)^2$, which separate the different regions in the framework of the previous model. It is seen that the boundaries between the different regions are changed due to the dependence of the critical current on the magnetization direction. Region III disappears in this model, and region IV is expanded. It is also seen from Fig.~\ref{s4}(a) that region IV can be divided into two subregions. Only "edge" non-aligned states with $m_{y1} = \pm 1$ or $m_{y2} = \pm1$ are possible in subregion IVa, analogously to the previous model. At the same time, additional non-aligned states, corresponding to $m_{y1} \neq \pm 1$ and $m_{y2} \neq \pm 1$ appear in subregion IVb. The reason is the suppression of the critical current by $m_x$. The lower critical current means the  smaller Josephson energy at a given phase difference, which is more energetically favorable. The suppression is also controlled by the parameter $r$, as it can be seen from Eq.~(\ref{critical_current}). Consequently, from the point of view of the Josephson energy it is favorable to enhance $m_x$. This tendency competes with the magnetic anisotropy energy, which tends to enhance $m_y$.  Therefore, at large enough values of $r$ and, simultaneously,  small enough $E_M/E_J$ the non-edge states  $m_{y1} \neq \pm 1$ and $m_{y2} \neq \pm 1$ can become energetically favorable, which is realized in region IVb.

The described above competition between the Josephson and magnetic energies is further illustrated in Fig.~\ref{s5}. It demonstrates the system energy as a function of $(m_{y1},m_{y2})$ for the same parameters, which are used for Figs.~\ref{config}(d) and (h). The only difference between the corresponding figures is that $I_c$ does not depend on $m_x$ in Fig.~\ref{config} and it depends on $m_x$ in Fig.~\ref{s5}. It is seen that at small $r=0.7$ the difference between the corresponding Figs.~\ref{config}(d) and \ref{s5}(a) is not essential. At the same time at $r=2.5$ Figs.~\ref{config}(h) and \ref{s5}(b) are qualitatively different. The reason is connected to the suppression of the critical current at nonzero $m_x$ and the resulting energy gain, as it is described above. 
\begin{figure*}[tbh!]
 \centerline{$
 \begin{array}{c}
 \includegraphics[width=5.5in]{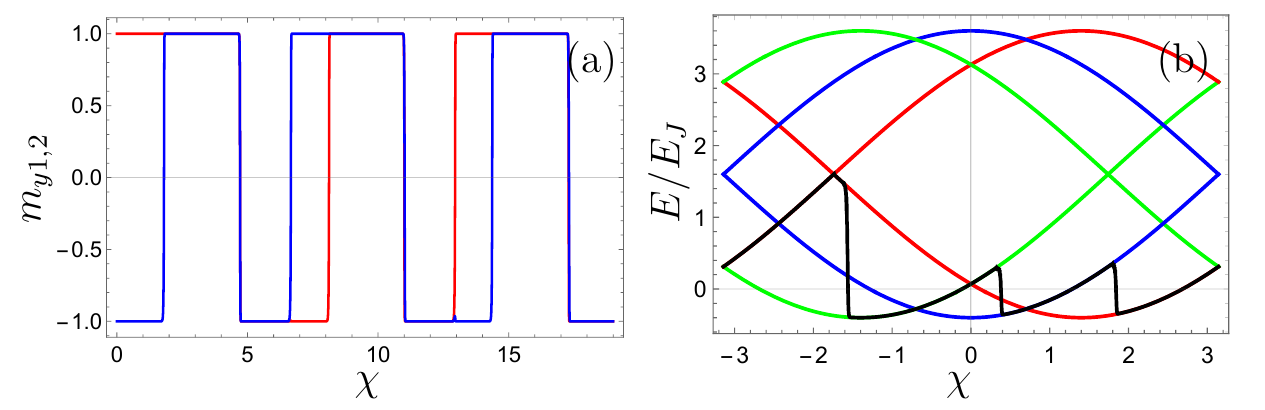} 
 \end{array}$}
 \caption{(a) Time evolution of $m_{y,1,2}$ under the adiabatic variation of the phase $\chi= 2eV t$ for the S/F/S JJ on top of the 3D TI. $eVt_0 =5\times 10^{-4}$. (b) Matching the dynamical magnetic configuration, presented in (a) to the energy of the system. Black line is the dynamical trajectory of the system. $r=0.7$, $E_M/E_J =0.2$.}
\label{s6} 
\end{figure*}

Further in Fig.~\ref{s6} we demonstrate the influence of $I_c(m_x)$ on the dynamics of the magnetic configuration under the adiabatic phase variations. This figure can hardly be differed from Fig.3(a)-(b). First of all, the energy branches of the corner states do not differ at all. It is natural because $m_x=0$ for the corner states and, therefore, the dependence $I_c(m_x)$ does not influence them. Moreover, the dynamical trajectory is also very similar. It is valid for small enough $r$, because in this case the non-aligned states are energetically close to the corner states and only exist in the narrow regions of the superconducting phase $\psi_0$. For this reason the system practically does not occur in the non-aligned states. At larger $r$ the regions of the non-aligned states existence expand and the dynamics can be modified. However, these regions probably are not of great interest for studying because of the strong Josephson current suppression at the magnetization orientations $m_x \neq 0$. The suppression strongly weakens the interaction between the magnets, mediated by the Josephson coupling. 
\bibliography{refs_LRM}

\end{document}